\documentclass[twocolumn,showpacs,preprintnumbers,amsmath,amssymb,aps]{revtex4-1}
\usepackage{graphicx}% Include figure files
\usepackage{dcolumn}% Align table columns on decimal point
\usepackage{bm}% bold math
\usepackage{float}
\usepackage{soul}

\begin{document}

\title{A precise measure of avian magnetoreception based on quantum metrology}

\author{Li-Sha Guo$^{1}$}
\author{Bao-Ming Xu$^{2}$}
\author{Jian Zou$^{1}$}%
\email{zoujian@bit.edu.cn}
\author{Bin Shao$^{1}$}%

\affiliation{$^{1}$School of Physics, Beijing Institute of Technology, Beijing 100081, China}%
\affiliation{$^{2}$School of Physics, Qufu Normal University, Qufu 273165, China}

\date{Submitted \today}

\begin{abstract}
The radical pair (RP) mechanism, which describes the quantum dynamics of a spatially separated electron pair,
is considered as one of the principal models of avian magnetoreception. Different from the conventional phenomenological approach where the sensitivity of avian magnetoreception is characterized by the singlet yield $\Phi_{S}$, we introduce the quantum Fisher information (QFI), which represents the maximum information about the magnetic field's direction extracted from the RP state, to give a precise measure of sensitivity of avian compass essentially. The consistency between our results and experimental observations suggests that QFI plays a decisive role in avian magnetoreception. Besides, within the framework of quantum metrology, we can judge the feasibility of any possible measurement scheme for avian magnetoreception, and shed light on an intrinsic relevance between the singlet yield and a concrete measurement scheme of our approach. The present work allows us to understand many things about avian magnetoreception from a fully new perspective of quantum  metrology, and provide a new route to establish a direct connection between quantum information and many other biological functions.

\end{abstract}

\pacs{03.67,-a, 06.20.-f, 03.65.Yz, 82.30.-b}

\maketitle

\textit{Introduction}.
Recent evidence suggests that some unique features of quantum mechanics can be harnessed to enhance biological functions in a large variety of living organisms, e.g., in natural selection \cite{Lloyd1}, olfaction sense \cite{Brookes,Solov'yov}, enzymatic reactions \cite{enzymatic1,enzymatic2}, photosynthetic light harvesting \cite{Plenio1,Lloyd2}, avian magnetoreception \cite{Schulten1978,Ritz2000,Maeda2008,Hore2009,Kominis,Xu2014,Xu2016,Gauger2011,Yang2012,Sun2012,Cai2010,Imamoglu2015,Cai2012,Hore2012,Cai2013,Jayendra,Tiersch2012,Pauls,Chia1,Mouloudakis}, etc., which indicates that quantum biology has been entering a new stage \cite{Nori,Vedral,Plenio2}. As one of the principal models of avian magnetoreception, the radical pair (RP) mechanism \cite{Schulten1978,Ritz2000,Hore2009}, based on singlet-triplet transitions due to the anisotropic hyperfine (HF) interaction, suggests that migratory birds depend on the photoinduced RPs for navigation, which has been supported
by intensive evidences and behavioral experiments with birds \cite{Wiltschko1972,Wiltschko1978,Wiltschko2006,Wiltschko2013,Thalau2005,Ritz2004,Ritz2009}.
Due to the quantum mechanical nature of RP model, a growing interest in understanding the function of avian magnetoreception
has extended from chemists, biologists to physicists, by using the rich fruits in the field of quantum information such as quantum coherence and entanglement \cite{Pauls,Cai2010,Cai2013,Hore2012,Jayendra,Cai2012,Gauger2011,Xu2016}. However, a deeper understanding of the mechanism of avian magnetoreception may need the ability to precisely measure the function of avian compass. With the development of various kinds of quantum techniques, particularly in the field of quantum metrology \cite{Lloyd3,Lloyd4,Taylor}, which has primarily been developed to find the fundamental limit to precision of estimating an unknown parameter, can we use the method of quantum metrology to precisely characterize the sensitivity of avian magnetoreception?

In this letter, we apply quantum metrology to avian magnetoreception, and use the quantum Fisher information (QFI) to give a precise measure of sensitivity of avian magnetoreception, which represents the maximum information about the geomagnetic field direction extracted from the RP state. Such an approach allows us to establish a quantitative connection between the performance of avian compass and the magnitude of QFI. Although there have been a few works which noticed the potential relevance between quantum metrology and biology \cite{Mouloudakis,Taylor,Cai2013}, we have not see any relevant works to date which really characterized the magnetic sensitivity of avian compass by means of quantum metrology. In the context of RP model, we first derive a statistical average state (i.e., a steady state) of RP, then calculate the QFI of this state and finally compare the results with the relevant experimental results. The highly consistency between the behavior of QFI and the experimental results underlies a decisive role played by the QFI in avian magnetoreception.
Besides, within the framework of quantum metrology, we can judge the feasibility of any possible measurement scheme for avian magnetoreception, and shed light on an intriguing connection between the conventional approach (i.e., the singlet yield) \cite{Schulten1978,Ritz2000} and a concrete measurement scheme of our approach.

\textit{RP Model and its statistical average state}. In the avian compass, each photoinduced RP has a spatially separated electron pair coupled to an external magnetic field $\textbf{B}$ and a few nuclei. Generally it is believed that only one of the electrons interacts with the nuclei with an anisotropic HF coupling and the other is free \cite{Maeda2008}. Thus this provides asymmetry and leads to singlet-triplet transition required for the directional sensitivity. In this letter, we only consider the simple case of one nuclear spin and the corresponding
Hamiltonian for each RP is \cite{Kominis,Xu2014,Cai2012,Gauger2011,Jayendra,Xu2016,Gauger2011}
\begin{equation}\label{H}
    H=\gamma \textbf{B}\cdot(\hat{S}_{1}+\hat{S}_{2})+\hat{I}\cdot \textbf{A}\cdot \hat{S}_{2},
\end{equation}
where $\mathbf{B}$ is the external magnetic field around the RP, $\gamma=\frac{1}{2}\mu_{B}g_{s}$ is the gyromagnetic ratio, with $\mu_{B}$ being the Bohr's magneton and $g_{s}$ being the $g$ factor of electron. Here, we assume that the $g$ factor is the same for both electronic spins and set its value according to free electron, i.e., $g_{s}=2$. $\hat{S}_{i}=(\sigma_{x},\sigma_{y},\sigma_{z})$ are the electronic spin operators ($i=1,2$), and $\hat{I}$ is the nuclear spin $1/2$ operator. $\textbf{A}$ is the HF tensor which couples the nuclear spin and electron 2 with a diagonal form $\mathbf{A}=diag(A_{x},A_{y},A_{z})$, and we assume an axially symmetric (or cigar-shaped) HF tensor, i.e., $A_{z}>A_{x}=A_{y}$.
The RP density matrix at time $t$ can then be described as
\begin{equation}\label{rho}
 \rho_{s}(t)=\mathrm{Tr}_{I}[U(t)\rho(0)U^{\dag}(t)],
\end{equation}
where $U(t)$ is the evolution operator corresponding to the Hamiltonian Eq. (\ref{H}), and $\mathrm{Tr}_{I}[\cdot]$ means taking the trace over the nucleus. $\rho(0)=\rho_{s}(0)\otimes\rho_{I}(0)$ is the initial state of two electrons and one nucleus, and generally the nucleus is initially in a complete mixed state, i.e., $\rho_{I}(0)=\mathbb{I}/2$.

First we assume that the RPs are identical and in the same initial state. Due to the continuous optical excitation, the creation of each RP is entirely accidental and its decay is also random. However, with respect to all the RPs existing in the bird's eye, they would be in a steady state. In what follows, we would derive a statistical average state of RP to describe this steady state. To be more specific, choosing an arbitrary fixed time to see (here we set the fixed time as the reference time, denoted as $t'=0$), the RPs at the reference time ($t'=0$) are constituted of those evolved from different time $t'$ ($t'<0$), i.e., the moment of RP formation. It is reasonable to assume that in time regime $t'$$\sim$$t'+dt'$, the number of RPs created by optical excitation is a constant which is not dependent on the specific time $t'$, denoted by $\Delta M$. And the number of them which still exist (not decay) at the reference time is $d\Delta M(t')=\Delta Mf(t')dt'$, where $f(t')\equiv k\exp(-k|t'|)$, with $k$ being the recombination rate \cite{Steiner1989}.
In other words, for each RP created by optical excitation in time regime $t'$$\sim$$t'+dt'$, its existing probability at the reference time is
\begin{equation}\label{Pt}
  P(t')=\frac{d\Delta M(t')}{\Delta M}=f(t')dt',
\end{equation}
and the corresponding state at the reference time ($t'=0$) is described as $\rho_{s}(t')$ which is evolved from the time regime $t'$$\sim$$t'+dt'$.
Due to the fact that each RP is subject to the optical excitation randomly, at the reference time, the state of the RP would be consisted of a large number of states evolved from different time $t'$ with a corresponding weight $P(t')$.
As a result, we can obtain a statistical average state (i.e., the steady state) of RP:
\begin{equation}\label{rhoba}
  \bar{\rho}_{s}=\int_{-\infty}^{0}f(t')\rho_{s}(t')dt'=\int_{0}^{\infty}f(t)\rho_{s}(t)dt,
\end{equation}
where in the second equation, we have replaced the integration variable $t'$ with $t=-t'$, and accordingly $\rho_{s}(t')$ is equal to $\rho_{s}(t)$ defined in Eq. (\ref{rho}).
Here it is noted that $\int_{-\infty}^{0}f(t')dt'=\int_{0}^{\infty}f(t)dt=1$.

\textit{Magnetic sensitivity quantified by QFI}.
For the avian compass, the estimated parameter is the geomagnetic field orientation to the basis of HF tensor. And according to the quantum parameter estimation theory (refer to Appendix A for a brief introduction), the QFI for estimating an unknown parameter $x$ can be obtained as \cite{Braunstein1994,H1999}
\begin{equation}\label{qqfi}
    \mathrm{QFI}=2\sum_{p_{j}+p_{k}\neq0}\frac{1}{p_{j}+p_{k}}\Big\vert\langle\psi_{j}|\frac{d\rho^{x}}{dx}|\psi_{k}\rangle\Big\vert^{2},
\end{equation}
where $\rho^{x}$ is the parameter dependent state, with $|\psi_{i}\rangle$ being its eigenstate and $p_{i}$ its corresponding eigenvalue. In what follows we would calculate the QFI of the steady state $\bar{\rho}_{s}$ of RP.
In the main text, we only consider a simple case where $A_{x} = A_{y} = 0$ (and the case $A_{x} = A_{y}\neq 0$ is considered in Appendix C). Generally, the geomagnetic field can be described as
\begin{equation}\label{B0}
  \mathbf{B_{0}}=\mathrm{B}_{\mathrm{0}}(\sin\theta\cos\phi,\sin\theta\sin\phi,\cos\theta),
\end{equation}
where $\mathrm{B}_{\mathrm{0}}$ is the intensity of the geomagnetic field, and $\theta$ and $\phi$ describe the orientation of the geomagnetic field to the basis of HF tensor. The axial symmetry of HF tensor allows us to set $\phi=0$ and focus on $\theta$ in the range $[0,\pi/2]$ without loss of generality, and $\theta$ is the parameter to be estimated for the avian compass.
And then we can calculate the QFI of the steady state $\bar{\rho}_{s}$ of RP under the influence of the geomagnetic field, and in this case, $\mathbf{B}=\mathbf{B_{0}}$ in Eq. (\ref{H}). In most previous studies of avian compass, the recombination rate $k$ is generally considered to be the order of $10^{4}s^{-1}\sim 10^{6}s^{-1}$. And in this regime, for an arbitrary initial state of RP $\rho_{s}(0)$, an approximate expression of QFI of the steady state $\bar{\rho}_{s}$ can be obtained, by making a strong HF coupling approximation, i.e., $A_{z}\gg \gamma \mathrm{B}_{\mathrm{0}}$ (the detailed derivation of QFI can be seen in Appendix B):
\begin{equation}\label{QFIa}
 \mathrm{QFI}\approx\sum_{i=0}^{1}\mathrm{Re}[\rho_{i}^{12}]^{2}(\frac{1}{\rho_{i}^{11}}+\frac{1}{\rho_{i}^{22}})+\frac{(\rho_{i}^{11}-\rho_{i}^{22})^{2}}{\rho_{i}^{11}+\rho_{i}^{22}},
\end{equation}
where $\rho^{ij}_{1}=\langle\phi_{i}|\langle1|\rho_{s}(0)|\phi_{j}\rangle|1\rangle$, and $\rho^{ij}_{0}=\langle\phi_{i}|\langle0|\rho_{s}(0)|\phi_{j}\rangle|0\rangle$,
with $|0\rangle$ ($|1\rangle$) and $|\phi_{i}\rangle$ ($i=1,2$) being the eigenstates of $\sigma_{z}$ of electron 2 and Hamiltonian of electron 1, i.e., $H_{1}=\gamma\mathbf{B_{0}}\cdot \hat{S}_{1}$, respectively, and $\mathrm{Re}[\rho_{i}^{12}]$ represents the real part of $\rho_{i}^{12}$. From Eq. (\ref{QFIa}) we can see that for any given initial state $\rho_{s}(0)$, the QFI of the steady state $\bar{\rho}_{s}$ is not dependent on $\mathrm{B}_{\mathrm{0}}$, which implies that the change of the intensity of external magnetic field $\mathrm{B}_{\mathrm{0}}$ would not disorient the bird permanently. This is consistent with the experimental result that bird can adapt to different magnetic field intensities \cite{Wiltschko1978,Wiltschko2006,Wiltschko2013}. Furthermore, without making any approximation, we numerically plot the QFI for different magnetic field intensities with the RP initial state being the singlet state $|S\rangle=\frac{1}{\sqrt{2}}(|10\rangle-|01\rangle)$ in Fig. 1 as an example. We can see from Fig. 1 that the $30\%$ weaker (32.2$\mu$T) and stronger (59.8$\mu$T) fields \cite{Wiltschko1978} compared with the geomagnetic field (46$\mu$T) have almost no influences on the value of QFI, that is, bird would not disorient when the intensity of magnetic field is decreased or increased by about $30\%$ of that of geomagnetic field. It is noted that for $k=10^{5}s^{-1}$ and $10^{6}s^{-1}$, the above results also hold.

\begin{center}
\includegraphics[width=6cm]{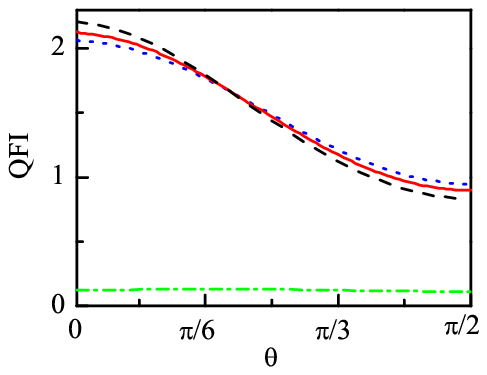}
\parbox{8cm}{\small{FIG. 1.} (Color online) The QFI as a function of the direction angle $\theta$ without the oscillating field ($\mathrm{B}_{\mathrm{0}}=46\mu$T (red solid line), $\mathrm{B}_{\mathrm{0}}=59.8\mu$T (black dashed line), and $\mathrm{B}_{\mathrm{0}}=32.2\mu$T (blue dotted line)), and with the oscillating field $\mathrm{B}_{\mathrm{rf}}=150$nT and $\mathrm{B}_{\mathrm{0}}=46\mu$T (green dash dotted line). $A_{z} = 6\gamma\times46\mu$T, $A_{x}=A_{y}=0$, $k=10^4s^{-1}$.}
\end{center}

Next, we would investigate the influence of an additional weak resonant oscillating field, and in this case, $\mathbf{B}=\mathbf{B_{0}}+\mathbf{B_{rf}}$ in Eq. (\ref{H}) with
\begin{equation}\label{Brf}
    \mathbf{B_{rf}}=\mathrm{B}_{\mathrm{rf}}\cos wt(\sin\alpha\cos\beta,\sin\alpha\sin\beta,\cos\alpha),
\end{equation}
where $\mathrm{B}_{\mathrm{rf}}$ is the strength of oscillating field with frequency $\omega=2 \gamma \mathrm{B}_{\mathrm{0}}$ being resonant with the free electron. $\alpha$ and $\beta$ represent the direction of oscillating field with respect to the basis of HF tensor. Due to the axial symmetry of HF tensor we set $\beta=0$. Firstly, we consider $\alpha=\theta+\pi/2$, i.e., the weak oscillating field is perpendicular to Earth's magnetic field. In this case, when $k$ is in the regime $10^{4}s^{-1}\sim 10^{6}s^{-1}$, for an arbitrary initial state of RP $\rho_{s}(0)$, we can also obtain an approximate expression of QFI of the steady state $\bar{\rho}_{s}$, by making a strong HF coupling approximation (see Appendix B for a detailed derivation):
\begin{equation}\label{QFIBRF}
\begin{split}
\mathrm{QFI}\approx\sum_{i=0}^{1}\frac{k^4\mathrm{Re}[\rho_{i}^{12}]^2}{(k^2+\Omega^2)^2}\bigl(\frac{1}{P_{i}^{11}}+\frac{1}{P_{i}^{22}}\bigr)+\frac{(P_{i}^{11}-P_{i}^{22})^2}{P_{i}^{11}+P_{i}^{22}},
\end{split}
\end{equation}
where $\Omega=\gamma \mathrm{B}_{\mathrm{rf}}$, $P_{i}^{jj}=\rho^{jj}_{i} +(-1)^j \chi_{i}$, with $\rho^{jj}_{i}$ having been defined below Eq. (\ref{QFIa}), $\chi_{i}=\frac{\Omega^2}{2(k^2 + \Omega^2)}(\rho_{i}^{11}-\rho_{i}^{22})-\frac{\Omega k}{(k^2 +\Omega^2)}\mathrm{Im}[\rho_{i}^{12}]$ ($i=0,1$,  $j=1,2$), and $\mathrm{Im}[\rho_{i}^{12}]$ represents the imaginary part of $\rho_{i}^{12}$. Through our calculation, we obtain that when $\Omega=0$ (without the oscillating field), Eq. (\ref{QFIBRF}) reduces to Eq. (\ref{QFIa}); when $\Omega\gg k$, QFI$\approx$0, which implies that when the order of $k$ is much smaller than that of $\gamma \mathrm{B}_{\mathrm{rf}}$, the weak resonant oscillating field can completely disorient the bird. It is noted that the present conclusions are also not dependent on the specific kind of initial state of RP.

Furthermore, without making any approximation, for $k=10^{4}s^{-1}$, we numerically plot the QFI with the weak resonant oscillating field orthogonal to the geomagnetic field in Fig. 1, for the RP initial state being the singlet state $|S\rangle$ as an example. We can see from Fig. 1 that the value of QFI is highly reduced when the weak resonant oscillating field is applied, which satisfies the experimental result that a weak resonant oscillating field can disrupt the bird completely \cite{Ritz2004,Thalau2005,Ritz2009}. Besides, we find that the oscillating field parallel to Earth's magnetic field does not affect the value of QFI which is consistent with the experimental results \cite{Ritz2004,Thalau2005,Ritz2009}. Here we emphasize that in Appendix B, we not only discuss the order of $k$ in terms of QFI, and obtain that $k$ should be the order of $10^{4}s^{-1}$, which is in accordance with the previous works \cite{Xu2014,Gauger2011,Yang2012}, but also numerically show that when $k=10^{4}s^{-1}$, for an arbitrary initial state of RP, a weak resonant oscillating field orthogonal to the geomagnetic field can reduce the value of QFI by at least $87\%$ of that without the oscillating field.

\textit{Discussion\textemdash possible implementations for avian compass}.
In this letter, we use the QFI to quantify the magnetic sensitivity of avian compass, but it is only an upper bound of precision for magnetoreception. In fact, there may exists several possible implementations, and what specific kind of implementation is adopted by birds in nature is not clear for us, despite of the prevailing view that the external magnetic field information can be recorded by the singlet yield \cite{Schulten1978,Ritz2000} which can be detected by birds. Given that the initial state of RP is in the singlet state $|S\rangle$, we give two possible implementations as examples here, which are the measurement of total angular momentum and that of the square of total magnetic moment.

When a specific POVM measurement, corresponding to the observable $\hat{A}$, has been performed, the unknown parameter $\theta$ can be estimated from the mean value of $\hat{A}$, with the precision given by the standard error propagation formula
$\Delta^{2}\theta=\frac{\Delta^{2} \hat{A}}{|d\langle \hat{A}\rangle/d\theta|^{2}}$ \cite{Helstrom1976,Holevo1982}, where $\Delta ^{2}\hat{A}$ and $\langle \hat{A}\rangle$ represent the variance and mean value of the observable $\hat{A}$ obtained for $\bar{\rho}_{s}$, respectively. Firstly, we give the measurement of total angular momentum, i.e., $\hat{A}=\hat{S}^{2}=(\hat{S}_{1}+\hat{S}_{2})^{2}$. Here it should be noted that $\langle \hat{S}^{2}\rangle=2(1-P_{S})$, with $P_{S}\equiv\langle S|\bar{\rho}_{s}|S\rangle$ representing the probability that the RP is found in the singlet state $|S\rangle$, besides, it can be seen from Eq. (\ref{rhoba}) that $\langle S|\bar{\rho}_{s}|S\rangle=\int_{0}^{\infty}f(t)\langle S|\rho_{s}(t)|S\rangle dt\equiv \Phi_{S}$ \cite{Xu2014,Cai2012,Gauger2011,Jayendra,Xu2016,Gauger2011}. Thus, the widely used signal contrast $D_{s}=\Phi_{max}-\Phi_{min}$ \cite{Cai2012,Hore2012,Cai2013,Jayendra}, which denotes the difference between the maximum and the minimum singlet yields along all the directions, is actually corresponding to the measurement of $\hat{S}^{2}$. Moreover, through our calculations, we find that $\Delta^{2}\theta$ is equal to the inverse of the classical Fisher information $1/\mathrm{F}$. Although the signal contrast $D_{s}$ and the classical Fisher information F(1/$\Delta^{2}\theta$) both describe the magnetic sensitivity of avian compass, F(1/$\Delta^{2}\theta$), which denotes the maximum information about $\theta$ extracted from the steady state of RP for this measurement scheme, is more accurate and can better reflect the essence of avian magnetoreception than $D_{s}$. Our numerical results of $1/\Delta^{2}\theta$ are shown in Fig. 2(a), and we can see that the $30\%$ stronger and weaker fields compared with Earth's magnetic field almost have no influences on the value of $1/\Delta^{2}\theta$, however, a weak resonant oscillating field perpendicular to Earth's magnetic field reduces the value of $1/\Delta^{2}\theta$ dramatically.

\begin{center}
\includegraphics[width=8cm]{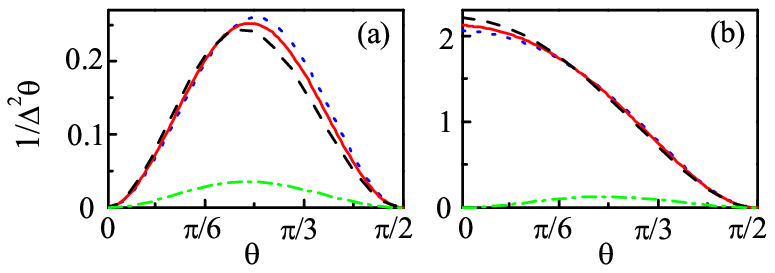}
\parbox{8cm}{\small{FIG. 2.} (Color online) $1/\Delta^{2}\theta$ as a function of the direction angle $\theta$ for measuring (a) $\hat{S}^{2}$ and (b) $\hat{S}_{z}^{2}$ without the oscillating field ($\mathrm{B}_{\mathrm{0}}=46\mu$T (red solid line), $\mathrm{B}_{\mathrm{0}}=59.8\mu$T (black dashed line), and $\mathrm{B}_{\mathrm{0}}=32.2\mu$T (blue dotted line)), and with the oscillating field $\mathrm{B}_{\mathrm{rf}}=150$nT and $\mathrm{B}_{\mathrm{0}}=46\mu$T (green dash dotted line). For both (a) and (b) the initial state of RP is the singlet state with $A_{z} = 6\gamma\times46\mu$T, $A_{x}=A_{y}=0$ and $k=10^{4}s^{-1}$.}
\end{center}

Next, we give the measurement of the square of magnetic moment, i.e., $\hat{A}=\hat{S}_{z}^{2}=(\hat{S}_{1z}+\hat{S}_{2z})^{2}$, where $\hat{S}_{1z}$ and $\hat{S}_{2z}$ represent the $z$ component of spin angular momentum of electron 1 and 2 of RP, respectively. When the initial state of RP is in the singlet state $|S\rangle$, $\langle \hat{S}_{z}\rangle=0$. In this case, the measurement of $\hat{S}_{z}^{2}$ can be considered as a sense of fluctuation of magnetic moment, for $\Delta^{2}\hat{S}_{z}=\langle\hat{S}_{z}^{2}\rangle-\langle\hat{S}_{z}\rangle^{2}=\langle \hat{S}_{z}^{2}\rangle$. It is noted that $\Delta^{2}\theta$ is also equal to the inverse of the classical Fisher information $1/\mathrm{F}$ through our calculations. Our numerical results of $1/\Delta^{2}\theta$ are shown in Fig. 2(b), and we can see that $1/\Delta^{2}\theta$ is robust to different magnetic field intensities, but would be highly reduced when a weak resonant oscillating field is applied. In fact, in the spirit of this line, we can judge the feasibility of any possible measurement scheme for avian magnetoreception by comparing the corresponding measurement results with the relevant experiment results.

\textit{Other results}.
Following the present insight that the QFI can well quantify the magnetic sensitivity of avian compass, it is possible to study the effects of entanglement and different decoherence models on the value of QFI in a unified picture. Here we also take the singlet state $|S\rangle$ as the initial state of RP as an example. Firstly, we find that for an arbitrary direction angle $\theta$, when $k$ is small, the QFI is relatively large while the entanglement is equal to 0, however, when $k$ is large, the entanglement becomes large while the QFI reduces to 0, which implies that entanglement can not help to promote bird orientation (a detailed discussion can be seen in Appendix D). Next, we investigate the effects of three typical classes of independent Markovian environmental noise on the value of QFI, namely, the amplitude damping noise, dephasing noise and depolarized noise. By comparing our numerical results with the experimental observations, we find that for the amplitude damping noise and the dephasing noise, the decoherence rate should be smaller than 10$k$, while for the depolarized noise, the decoherence rate should even be smaller than $k$ (a detailed discussion can be seen in Appendix E).

\textit{Summary and outlook}. As a precise measure of avian compass, the QFI essentially determines the ability of migratory birds to sense the direction of Earth's magnetic field. Compared with the conventional phenomenological approach where the magnetic sensitivity is quantified by the signal contrast $D_{s}=\Phi_{max}-\Phi_{min}$, our approach proves more accurate and can better reflect the essence of avian magnetoreception. Meanwhile, in this unified approach of QFI, the order of the recombination rate and the effects of entanglement and decoherence on avian magnetoreception can be well understood. Considering that the QFI is only an upper bound of precision for directional detection, it is desirable to seek for a potential measurement scheme to characterize the compass sensitivity. In the spirit of our approach, we can judge the feasibility of any possible measurement scheme for avian magnetoreception, and have found that the conventional phenomenological approach, i.e, the singlet yield, is just one of the several feasible measurement schemes. Then an open question naturally arises: Is the widely used singlet yield the optimal choice for describing the sensitivity of avian compass in a practical way? We hope that our approach can provide a new route to apply the QFI into many other biological processes, such that a precise measure of biological function can be given, and a more profound understanding of biological phenomena can be obtained, which may in turn give us a few clues in the quest to develop quantum technology.

\textit{Acknowledgements}. This work was supported by the National Natural Science
Foundation of China (Grants No. 11274043, 11375025).

L. S. Guo and B. M. Xu contributed equally to this work.

\appendix

\section{quantum parameter estimation theory}
A standard scenario in quantum parameter estimation can be described as follows: Firstly, a probe system would be prepared in an appropriate initial state $\rho(0)$, and then it undergoes an evolution which would imprint the parameter information onto the evolved state, say $\rho^{x}$, and finally it would subject to a POVM measurement. The overall process is repeated $\nu$ times, and we infer the parameter $x$ from the statistics of the measurement outcomes by choosing an unbiased estimator. The variance of this estimator, i.e., $\Delta^{2}x$, quantifies the error on estimation of $x$, and is
lower bounded by:
\begin{equation}\label{abc}
 \Delta^{2} x\geq\frac{1}{\nu \mathrm{F}}\geq\frac{1}{\nu \mathrm{QFI}},
\end{equation}
where $\mathrm{F}$ is the classical Fisher information optimized over all the possible estimators, and QFI is the quantum Fisher information, which is further optimized over all the allowable measurements and is given by \cite{Helstrom1976,Holevo1982,Braunstein1994}
\begin{equation}\label{qfi}
    \mathrm{QFI}=\mathrm{Tr}\bigl[\rho^{x} L^{2}_{\rho^{x}}\bigr], 
\end{equation}
where the symmetric logarithmic derivative $L_{\rho^{x}}$ in the above equation is defined as:
\begin{equation}\label{L}
    \frac{d\rho^{x}}{dx}\equiv\frac{1}{2}\bigl(\rho^{x} L_{\rho^{x}}+L_{\rho^{x}}\rho^{x}\bigr). 
\end{equation}
Writing $\rho^{x}$ in its spectral decomposition as $\rho^{x}=\sum_{i}p_{i}|\psi_{i}\rangle\langle\psi_{i}|$, one can obtain \cite{Braunstein1994}:
\begin{equation}\label{qqfi}
    \mathrm{QFI}=2\sum_{p_{j}+p_{k}\neq0}\frac{1}{p_{j}+p_{k}}\Big\vert\langle\psi_{j}|\frac{d\rho^{x}}{dx}|\psi_{k}\rangle\Big\vert^{2}. 
\end{equation}

\section{derivation of QFI}
In this section, we would derive the approximate expressions of QFI for an arbitrary initial state of RP with and without the oscillating field, i.e., Eq. (7) and Eq. (9) in the main text, respectively. When the horizontal HF coupling $A_{x} = A_{y} = 0$, the role of nuclear spin can be considered as applying an effective magnetic field (depending on its state) on the electronic spin. If the nucleus is in the spin up (down) state, the effective magnetic field is $A_{z}\hat{z}/\gamma (-A_{z}\hat{z}/\gamma)$, with $\hat{z}$ being the $z$ direction. As a result, the effective Hamiltonian of RP can be written as $H_{\pm}=\gamma\mathbf{B_0}\cdot(\hat{S}_{1}+\hat{S}_{2})\pm A_{z}\hat{S}_{2z}$, where $\mathbf{B_0}=\mathrm{B}_{\mathrm{0}}(\sin\theta\cos\phi,\sin\theta\sin\phi,\cos\theta)$ is the geomagnetic field around the RP, with $\mathrm{B}_{\mathrm{0}}$ being the intensity of the geomagnetic field, and $\theta$ and $\phi$ being the orientation of the geomagnetic field to the basis of the HF tensor. The axial symmetry of the HF tensor allows us to set $\phi=0$ and focus on $\theta$ in the range $[0,\pi/2]$ without loss of generality, and $\theta$ is the parameter to be estimated for avian compass. Here we denote the eigenstates of the effective Hamiltonian $H_{\pm}$ as $|\Psi^{i}_{\pm}\rangle\in\{|\phi_{1}\rangle|\psi_{1\pm}\rangle,|\phi_{1}\rangle|\psi_{2\pm}\rangle,|\phi_{2}\rangle|\psi_{1\pm}\rangle,|\phi_{2}\rangle|\psi_{2\pm}\rangle\}$ and its corresponding eigenvalues as $E_{\pm}^{i}$ ($i=1,2,3,4$). Specifically, $|\phi\rangle_{1}=\cos\frac{\theta}{2}|1\rangle+\sin\frac{\theta}{2}|0\rangle$ and $|\phi\rangle_{2}=\sin\frac{\theta}{2}|1\rangle-\cos\frac{\theta}{2}|0\rangle$ are the eigenstates of Hamiltonian of electron 1, i.e., $H_{1}=\gamma\mathbf{B_0}\cdot \hat{S}_{1}$. $|\psi_{1\pm}\rangle=\cos\frac{\theta_{\pm}}{2}|1\rangle+\sin\frac{\theta_{\pm}}{2}|0\rangle$ and $|\psi_{2\pm}\rangle=\sin\frac{\theta_{\pm}}{2}|1\rangle-\cos\frac{\theta_{\pm}}{2}|0\rangle$ are the eigenstates of Hamiltonian of electron 2, i.e., $H_{2\pm}=\gamma\mathbf{B_0}\cdot\hat{S}_{2}\pm A_{z}\hat{S}_{2z}$, with $\sin\theta_{\pm}=B_{x}/B_{\pm}$, $B_{x}=\mathrm{B_{0}}\sin\theta$, $\cos\theta_{\pm}=(B_{z}\pm A_{z}/\gamma)/B_{\pm}$, $B_{z}=\mathrm{B_{0}}\cos\theta$, and $B_{\pm}=\sqrt{B_{x}^2+(B_{z}\pm A_{z}/\gamma)^2}$ \cite{Cai2012,Xu2014}.

\subsection{Derivation of QFI without oscillating field}
Given an arbitrary initial state of RP $\rho_{s}(0)$, we would derive the approximate expression of QFI of the steady state $\bar{\rho}_{s}$ of RP (see Eq. (4) in the main text) without considering the oscillating field. We can always expand $\rho_{s}(0)$ in the eigenbasis of the effective Hamiltonian $H_{\pm}$ as
\begin{equation}\label{rho0}
  \rho_{s}(0)=\sum_{i,j=1}^{4} \rho^{ij}_{\pm}(0)|\Psi^{i}_{\pm}\rangle\langle\Psi^{j}_{\pm}|,  
\end{equation}
with $\rho^{ij}_{\pm}(0)=\langle\Psi^{i}_{\pm}|\rho_{s}(0)|\Psi^{j}_{\pm}\rangle$.
Generally, the nucleus is initially in a complete mixed state, i.e., $\rho_{I}(0)=\mathbb{I}/2$. As a result, the state dependent effective magnetic field $A_{z}\hat{z}/\gamma (-A_{z}\hat{z}/\gamma)$ induced by the nuclear spin leads to the effective Hamiltonian of RP $H_{+}(H_{-})$ with the same probability $1/2$.
After some calculations, we can obtain
the RP density matrix at time $t$ analytically:
\begin{equation}\label{rhot}
  \rho_{s}(t)=\frac{1}{2}\bigl(\rho_{+}(t)+\rho_{-}(t)\bigr) 
\end{equation}
with
\begin{equation}\label{rho3}
\rho_{\pm}(t)=\sum_{i,j=1}^{4} \rho^{ij}_{\pm}(0)e^{-i(E_{\pm}^{i}-E_{\pm}^{j})t}|\Psi^{i}_{\pm}\rangle\langle\Psi^{j}_{\pm}|. 
\end{equation}
In most previous studies of avian compass, the recombination rate $k$ is generally considered to be the order of $10^{4}s^{-1}\sim 10^{6}s^{-1}$. And in this regime, $E_{\pm}^{i} (\sim 10^{8}s^{-1})\gg k$, thus the high-frequency oscillating terms of Eq. (\ref{rhot}) have no contribution to the time integral of Eq. (4) in the main text, hence the steady state of RP can be expressed as
\begin{equation}\label{rhobar1}
  \bar{\rho}_{s}\approx\frac{1}{2}\sum_{i=1}^{4}\rho^{ii}_{+}(0)|\Psi^{i}_{+}\rangle\langle\Psi^{i}_{+}|+\rho^{ii}_{-}(0)|\Psi^{i}_{-}\rangle\langle\Psi^{i}_{-}|. 
\end{equation}
Now we consider the strong HF coupling approximation, i.e., $A_{z}\gg\gamma \mathrm{B}_{\mathrm{0}}$, and expand the eigenvectors $|\psi_{1\pm}\rangle$ and $|\psi_{2\pm}\rangle$ in a power series of $\gamma \mathrm{B}_{\mathrm{0}}/A_{z}$, keeping terms to the first order. Through our calculation, we obtain that $|\psi_{1\pm}\rangle\approx\frac{\gamma \mathrm{B}_{\mathrm{0}}}{2A_{z}}\sin\theta|1\rangle\mp|0\rangle$ and $|\psi_{2\pm}\rangle\approx|1\rangle\pm\frac{\gamma \mathrm{B}_{\mathrm{0}}}{2A_{z}}\sin\theta|0\rangle$.
Submitting them into Eq. (\ref{rhobar1}) and keeping terms to the first order of $\gamma \mathrm{B}_{\mathrm{0}}/A_{z}$, $ \bar{\rho}_{s}$ can be approximately simplified as a diagonal form:
\begin{equation}\label{rhobar2}
  \bar{\rho}_{s}\approx\sum_{i=1}^{2}\rho^{ii}_{1}|\phi_{i}\rangle\langle\phi_{i}|\otimes|1\rangle\langle1|+\rho^{ii}_{0}|\phi_{i}\rangle\langle\phi_{i}|\otimes|0\rangle\langle0| 
\end{equation}
with $\rho^{ij}_{1}=\langle\phi_{i}|\langle1|\rho_{s}(0)|\phi_{j}\rangle|1\rangle$, and $\rho^{ij}_{0}=\langle\phi_{i}|\langle0|\rho_{s}(0)|\phi_{j}\rangle|0\rangle$. And then according to Eq. (\ref{qqfi}), the QFI of $ \bar{\rho}_{s}$ (Eq. (\ref{rhobar2})) can be obtained analytically:
\begin{equation}\label{QFIa}
 \mathrm{QFI}\approx\sum_{i=0}^{1}\mathrm{Re}[\rho_{i}^{12}]^{2}\bigl(\frac{1}{\rho_{i}^{11}}+\frac{1}{\rho_{i}^{22}}\bigr)
 +\frac{(\rho_{i}^{11}-\rho_{i}^{22})^{2}}{\rho_{i}^{11}+\rho_{i}^{22}}, 
\end{equation}
where $\mathrm{Re}[\rho_{i}^{12}]$ represents the real part of $\rho_{i}^{12}$.

\subsection{Derivation of QFI with oscillating field}
Now we would derive the approximate expression of QFI of the steady state $\bar{\rho}_{s}$ (see Eq. (4) in the main text) for an arbitrary initial state of RP with a weak resonant oscillating field $ \mathbf{B_{rf}}=\mathrm{B}_{\mathrm{rf}}\cos wt(\sin\alpha\cos\beta,\sin\alpha\sin\beta,\cos\alpha)$, where $\mathrm{B}_{\mathrm{rf}}$ is the strength of oscillating field with frequency $\omega=2 \gamma \mathrm{B}_{\mathrm{0}}$ being resonant with electron 1. $\alpha$ and $\beta$ represent the direction of oscillating field with respect to the basis of the HF tensor. Due to the axial symmetry of the HF tensor we set $\beta=0$. Here we consider $\alpha=\theta+\pi/2$, namely, the weak oscillating field is perpendicular to Earth's magnetic field.
For the convenience of our calculation below, we express an arbitrary initial state of RP $\rho_{s}(0)$ as
\begin{equation}\label{}
  \rho_{s}(0)=\sum_{i,j=1}^{2} \varrho_{\pm}^{ij}(0)\otimes|\psi_{i\pm}\rangle\langle\psi_{j\pm}|
\end{equation}
with $\varrho_{\pm}^{ij}(0)=\langle\psi_{i\pm}|\rho_{s}(0)|\psi_{j\pm}\rangle$ representing the operator of electron 1. Because of the effect of nucleus, the Larmor frequency of electron 2 induced by the effective magnetic field and the geomagnetic field is always not resonant with the frequency of oscillating field, as a consequence, electron 2 can be considered as almost not influenced by the oscillating field \cite{Xu2014}. Based on this, the RP density matrix at time $t$ can be obtained as
\begin{equation}\label{rhott}
  \rho_{s}(t)=\frac{1}{2}\bigl(\rho_{+}(t)+\rho_{-}(t)\bigr)
\end{equation}
with
\begin{equation}\label{}
\rho_{\pm}(t)\approx\sum_{i,j=1}^{2} U(t)\varrho_{\pm}^{ij}(0) U^{\dag}(t)\otimes e^{-i(\varepsilon_{\pm}^{i}-\varepsilon_{\pm}^{j})t}|\psi_{i\pm}\rangle\langle\psi_{j\pm}|,
\end{equation}
where $\varepsilon_{\pm}^{i}=(-1)^{i+1}\gamma B_{\pm}$ are the eigenvalues of $H_{2\pm}=\gamma\mathbf{B_0}\cdot\hat{S}_{2}\pm A_{z}\hat{S}_{2z}$, and $U(t)=\overleftarrow{\mathrm{T}}\exp[-i\int_{0}^{t}\mathbb{H}_{1}(\tau)d\tau]$ is the evolution operator of electron 1 with $\mathbb{H}_{1}(t)=\gamma(\mathbf{B_{0}}+\mathbf{B_{rf}})\cdot\hat{S}_{1}$, and $\overleftarrow{\mathrm{T}}$ denoting the chronological time-ordering operator.
After performing the rotating-wave approximation, the evolution operator can be obtained in the eigenbasis $|\phi_{i}\rangle$ ($i=1,2$) of $H_{1}=\gamma\mathbf{B_{0}}\cdot \hat{S}_{1}$ \cite{Scully}:
\begin{equation}
\begin{split}
 U(t)=\begin{pmatrix}
 \cos\frac{\Omega t}{2}e^{-i\omega_{0} t} & i\sin\frac{\Omega t}{2}e^{-i\omega_{0} t} \\
 \\
 i\sin\frac{\Omega t}{2}e^{i\omega_{0} t} & \cos\frac{\Omega t}{2}e^{i\omega_{0} t}
\end{pmatrix},
\end{split}
\end{equation}
with $\omega_{0}=\gamma \mathrm{B_{0}}$, and $\Omega=\gamma\mathrm{B}_{\mathrm{rf}}$. When $k$ is in the regime $10^{4}s^{-1}\sim 10^{6}s^{-1}$,
$\varepsilon_{\pm}^{i}\gg k$, $\omega_{0}\gg k$, thus the high-frequency oscillating terms of Eq. (\ref{rhott}) have no contribution to the time integral of Eq. (4) in the main text, hence the steady state of RP under the influence of oscillating field can be expressed as
\begin{equation}\label{rr}
  \bar{\rho}_{s}\approx\frac{1}{2}(\bar{\rho}_{+}+\bar{\rho}_{-}), 
\end{equation}
with
\begin{equation}\label{}
\begin{split}
  \bar{\rho}_{\pm}&=P_{1\pm}|\phi_{1}\rangle\langle\phi_{1}|\otimes|\psi_{1\pm}\rangle\langle\psi_{1\pm}| \\
  &+P_{2\pm}|\phi_{1}\rangle\langle\phi_{1}|\otimes|\psi_{2\pm}\rangle\langle\psi_{2\pm}| \\
  &+P_{3\pm}|\phi_{2}\rangle\langle\phi_{2}|\otimes|\psi_{1\pm}\rangle\langle\psi_{1\pm}|  \\
  &+P_{4\pm}|\phi_{2}\rangle\langle\phi_{2}|\otimes|\psi_{2\pm}\rangle\langle\psi_{2\pm}|,
\end{split}
\end{equation}
where
\begin{equation}\label{}
\begin{split}
  P_{1\pm}&=\varrho_{\pm}^{11}(1,1)+\frac{\Omega k}{(k^2 + \Omega^2)}\mathrm{Im}[\varrho_{\pm}^{11}(1,2)] \\
  &- \frac{\Omega^2}{2(k^2 + \Omega^2)}\bigl(\varrho_{\pm}^{11}(1,1) - \varrho_{\pm}^{11}(2,2)\bigr),
\end{split}
\end{equation}
\begin{equation}\label{}
\begin{split}
P_{2\pm}&=\varrho_{\pm}^{22}(1,1) + \frac{\Omega k}{(k^2 + \Omega^2)}\mathrm{Im}[\varrho_{\pm}^{22}(1,2)] \\
  &- \frac{\Omega^2}{2(k^2 + \Omega^2)}\bigl(\varrho_{\pm}^{22}(1,1) - \varrho_{\pm}^{22}(2,2)\bigr),
\end{split}
\end{equation}
\begin{equation}\label{}
\begin{split}
P_{3\pm}&=\varrho_{\pm}^{11}(2,2) -\frac{\Omega k}{(k^2 + \Omega^2)} \mathrm{Im}[\varrho_{\pm}^{11}(1,2)] \\
  &+ \frac{\Omega^2}{2(k^2 + \Omega^2)}\bigl(\varrho_{\pm}^{11}(1,1) -\varrho_{\pm}^{11}(2,2)\bigr),
\end{split}
\end{equation}
\begin{equation}\label{}
\begin{split}
P_{4\pm}&=\varrho_{\pm}^{22}(2,2)- \frac{\Omega k}{(k^2 + \Omega^2)} \mathrm{Im}[\varrho_{\pm}^{22}(1,2)] \\
  &+ \frac{\Omega^2}{2(k^2 + \Omega^2)}\bigl(\varrho_{\pm}^{22}(1,1) - \varrho_{\pm}^{22}(2,2)\bigr)
\end{split} 
\end{equation}
with $\varrho_{\pm}^{ii}(m,n)=\langle\phi_{m}|\varrho_{\pm}^{ii}(0)|\phi_{n}\rangle$, ($i,m,n=1,2$).
Considering the strong HF coupling approximation, i.e., $A_{z}\gg\gamma \mathrm{B}_{\mathrm{0}}$, Eq. (\ref{rr}) can be approximately simplified
as a diagonal form:
\begin{equation}\label{ab}
\bar{\rho}_{s}\approx \sum_{i=1}^{2}P^{ii}_{1}|\phi_{i}\rangle\langle\phi_{i}|\otimes|1\rangle\langle1|+P^{ii}_{0}|\phi_{i}\rangle\langle\phi_{i}|\otimes|0\rangle\langle0|,
\end{equation}
where
\begin{equation}\label{}
  P_{i}^{jj}=\rho^{jj}_{i} +(-1)^j \chi_{i} 
\end{equation}
with $\rho^{jj}_{i}$ having been defined below Eq. (\ref{rhobar2}), $\chi_{i}=\frac{\Omega^2}{2(k^2 + \Omega^2)}(\rho_{i}^{11}-\rho_{i}^{22})-\frac{\Omega k}{(k^2 +\Omega^2)}\mathrm{Im}[\rho_{i}^{12}]$ ($i=0,1$,  $j=1,2$), and $\mathrm{Im}[\rho_{i}^{12}]$ represents the imaginary part of $\rho_{i}^{12}$.
And then according to Eq. (\ref{qqfi}), the QFI of $ \bar{\rho}_{s}$ (Eq. (\ref{ab})) can be obtained analytically:
\begin{equation}\label{qqfii}
\mathrm{QFI}\approx\sum_{i=0}^{1}\frac{k^4\mathrm{Re}[\rho_{i}^{12}]^2}{(k^2+\Omega^2)^2}\bigl(\frac{1}{P_{i}^{11}}+\frac{1}{P_{i}^{22}}\bigr)+\frac{(P_{i}^{11}-P_{i}^{22})^2}{P_{i}^{11}+P_{i}^{22}}. \end{equation}
Through our calculation, we obtain that when $\Omega=0$ (without the oscillating field), Eq. (\ref{qqfii}) reduces to Eq. (\ref{QFIa}); when $\Omega\gg k$, QFI$\approx$0, which implies that when the order of $k$ is much smaller than that of $\gamma \mathrm{B}_{\mathrm{rf}}$, the weak resonant oscillating field can completely disorient the bird.

\begin{center}
\includegraphics[width=8cm]{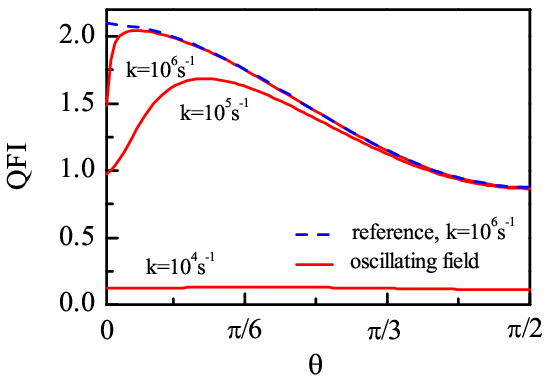}
\parbox{8cm}{\small{FIG. A1.} (Color online) The QFI as a function of direction angle $\theta$ with a weak resonant oscillating field perpendicular to Earth's magnetic field. $A_{z} = 6\gamma\times46\mu$T and $A_{x}=A_{y}=0$. The blue dashed line provides a reference of QFI without the oscillating field for $\mathrm{B}_{\mathrm{0}}=46\mu$T (The reference is independent of the recombination rate $k$ when $k\leq10^{7}s^{-1}$). The red solid lines represent the QFI when a 150nT resonant oscillating field is applied.}
\end{center}

\begin{center}
\includegraphics[width=8cm]{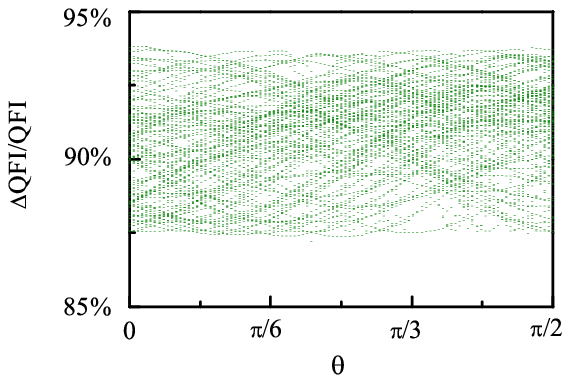}
\parbox{8cm}{\small{FIG. A2.} (Color online) The percent decrease of QFI when a weak resonant oscillating field orthogonal to the
geomagnetic field is applied, i.e., $\Delta\mathrm{QFI}/\mathrm{QFI}$, as a function of the direction angle $\theta$ for 100 randomly sampled initial states of RP with $A_{z} = 6\gamma\times46\mu$T, $A_{x}=A_{y}=0$, $\mathrm{B}_{\mathrm{0}}=46\mu$T, $\mathrm{B}_{\mathrm{rf}}=150$nT and $k=10^{4}s^{-1}$.}
\end{center}

In fact, the order of $k$ has been widely accepted to be approximately $10^{4}s^{-1}$$\sim$$10^{6}s^{-1}$ as mentioned above. In what follows, without making any approximation, we reconsider the order of $k$ in terms of QFI, by considering a weak resonant oscillating field of strength Brf=150nT perpendicular to Earth's magnetic field, which can completely disorient the bird \cite{Ritz2009}. Here we also take the singlet state $|S\rangle$ as the initial state of RP as an example. Our numerical results are shown in Fig. A1, and it can be seen that when $k=10^{6}s^{-1}$, the QFI is almost immune to the oscillating field, and when $k=10^{5}s^{-1}$, the QFI with the oscillating field is reduced to some extent compared with that without the oscillating field, but we are not sure whether this reduction of QFI can disrupt the birds or not. However, when $k=10^{4}s^{-1}$, the QFI with the oscillating field reduces significantly. Thus it is safe to say that if the oscillating field is to disorient the bird, it might be approximately $k=10^{4}s^{-1}$, which is consistent with the previous works \cite{Xu2014,Gauger2011,Yang2012}.

Below, we would further numerically show that when $k=10^{4}s^{-1}$, for an arbitrary initial state of RP, a weak resonant oscillating field orthogonal to the geomagnetic field can highly reduce the value of QFI of the steady state $\bar{\rho}_{s}$. Specifically, we randomly sample 100 initial states and plot in Fig. A2 the corresponding percent decreases of QFI, i.e., $\Delta\mathrm{QFI}/\mathrm{QFI}\equiv\frac{\mathrm{QFI}(\mathrm{B_{rf}=0})-\mathrm{QFI}(\mathrm{B_{rf}=150nT})}{\mathrm{QFI}(\mathrm{B_{rf}=0})}$, as a function of $\theta$. The results without making any approximation show that for all the sampled initial states, $\Delta\mathrm{QFI}/\mathrm{QFI}$ is larger than $87\%$, which implies that a weak resonant oscillating field orthogonal to the geomagnetic field can completely disorient the bird for an arbitrary initial state of RP.

\begin{center}
\includegraphics[width=8cm]{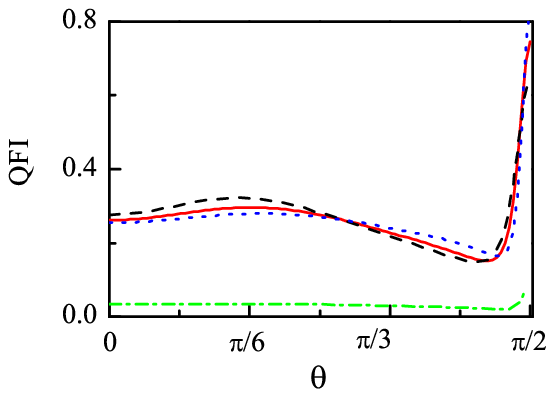}
\parbox{8cm}{\small{FIG. A3.} (Color online) The QFI as a function of the direction angle $\theta$ without the oscillating field ($\mathrm{B}_{\mathrm{0}}=46\mu$T (red solid line), $\mathrm{B}_{\mathrm{0}}=59.8\mu$T (black dashed line), and $\mathrm{B}_{\mathrm{0}}=32.2\mu$T (blue dotted line)), and with the oscillating field $\mathrm{B}_{\mathrm{rf}}=150$nT and $\mathrm{B}_{\mathrm{0}}=46\mu$T (green dash dotted line). $A_{z} = 6\gamma\times46\mu$T, $A_{x}=A_{y}=A_{z}/2$, $k=10^4s^{-1}$.}
\end{center}

\section{quantum fisher information with horizontal HF coupling}
The effect of external magnetic field and oscillating field on the value of QFI for $A_{x}=A_{y}=0$ has been discussed in the main text. Here we consider the case $A_{x}=A_{y}\neq 0$, and calculate the corresponding QFI. In this case, an approximately analytical expression of QFI can not be obtained, thus we calculate the QFI numerically. Through our large numerical calculations, we find that our results are not quite sensitive to what the value of HF coupling is. Here we consider $A_{z} = 6\gamma\times46\mu$T, $A_{x}=A_{y}=A_{z}/2$ and the initial state of RP to be the singlet state $|S\rangle$ as an example. The numerical results are shown in Fig. A3, and we can see that the QFI of $30\%$ weaker ($32.2 \mu$T) and stronger ($59.8 \mu$T) fields do almost not change compared with that of geomagnetic field ($46 \mu$T), but is highly reduced when a weak resonant oscillating field perpendicular to Earth's magnetic field is applied. Besides, through our numerical calculations, we find that there is no effect at such weak fields when the oscillating field is parallel to Earth's magnetic field. These results are similar to that without considering the horizontal HF coupling components in the main text.

\begin{center}
\includegraphics[width=8cm]{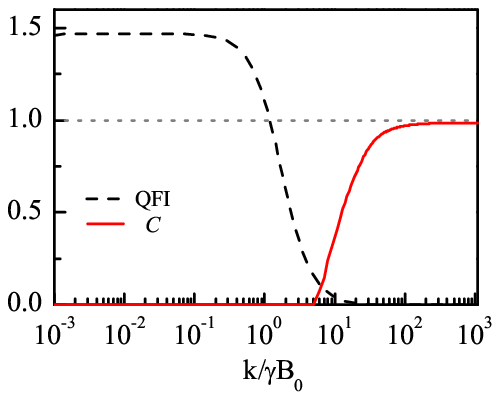}
\parbox{8cm}{\small{FIG. A4.} (Color online) The QFI (black dashed line) and concurrence $\textit{C}$ (red solid line) as a function of the recombination rate $k$ for $\mathrm{B}_{\mathrm{0}}=46\mu$T and $\theta=\pi/4$ with $A_{z} = 6\gamma\times46\mu$T, $A_{x}=A_{y}=0$.}
\end{center}

\section{effect of entanglement}
Due to the quantum mechanical nature of RP mechanism, the effect of entanglement on the avian compass has been investigated in terms of the singlet yield \cite{Gauger2011,Xu2016,Cai2010,Hore2012,Pauls,Zhang}. In this letter, we reconsider the effect of entanglement on the avian compass in terms of QFI, and use concurrence \cite{Wootters} to quantify entanglement. The concurrence $C$ of two qubits is defined as $C(\bar{\rho}_{s})=max\{0,~\sqrt{\lambda_{1}-\lambda_{2}-\lambda_{3}-\lambda_{4}}\}$, where $\lambda_{i}$ are the eigenvalues of the matrix $\bar{\rho}_{s}\sigma_{y}\otimes \sigma_{y}\bar{\rho}_{s}^{\ast}\sigma_{y}\otimes \sigma_{y} $ arranged in decreasing order, and $\bar{\rho}_{s}$ is defined in Eq. (4) in the main text. As an example, we plot the QFI and $C(\bar{\rho}_{s})$ as functions of the recombination rate $k$ for $\mathrm{B}_{\mathrm{0}}=46\mu$T, $\theta=\pi/4$, $A_{z} = 6\gamma\times46\mu$T, and $A_{x}=A_{y}=0$ with the RP initial state being the singlet state $|S\rangle$ in Fig. A4. And we can see from Fig. A4 that entanglement can not help to promote bird orientation, to be more specific, when $k$ is smaller, the QFI is relatively larger which actually corresponds to zero entanglement, and when $k$ is larger, the QFI is reduced to zero which corresponds to a relatively larger entanglement instead. It is noted that similar conclusions can be obtained for any other direction angles through our large numerical calculations. In fact, the behavior of entanglement as a function of $k$ can be also seen from the expression of $\bar{\rho}_{s}$. Specifically, when $k$ is small, $\bar{\rho}_{s}$ becomes a separable state (see Eq. (\ref{rhobar1})), which implies that there is no entanglement in $\bar{\rho}_{s}$ as shown in Fig. A4. However, when $k$ is large, it can be derived from Eq. (\ref{rhot}) and Eq. (\ref{rho3}) that $\bar{\rho}_{s}$ becomes the singlet state, because in this case the lifetime of RP ($\sim1/k$) is too short to make a transition between the singlet and triplet states, so the concurrence $C(\bar{\rho}_{s})$ equals to 1 as shown in Fig. A4.

\section{effect of decoherence}
Decoherence is unavoidable for the RP, and now we reconsider its effect on the avian compass in terms of QFI, with the singlet state $|S\rangle$ being the initial state of RP as an example. Specifically, we display three typical classes of independent Markovian environmental noise, namely, the amplitude damping noise, dephasing noise and depolarized noise.
We describe the environmental noises by the standard Lindblad master equation:
\begin{equation}\label{ME}
    \dot{\rho}(t)=-i[H,\rho(t)]+\sum_{i}\Gamma_{i}\bigl(L_{i}\rho(t)L_{i}^{\dag}-\frac{1}{2}\{L_{i}^{\dag}L_{i},\rho(t)\}\bigr),
\end{equation}
where $H=\gamma \textbf{B}\cdot(\hat{S}_{1}+\hat{S}_{2})+\hat{I}\cdot \textbf{A}\cdot \hat{S}_{2}$ (see Eq. (1) in the main text) denotes the total Hamiltonian for each RP, $\rho(t)$ represents the density matrix of one nucleus and two electrons at time $t$, $\Gamma_{i}$ represents the decoherence rate, $\{\cdot,\cdot\}$ represents the anticommutator, and $L_{i}$ is the Lindblad operator. 
For the amplitude damping noise, $L_{i}$ is only $\sigma_{-}$ for each electronic spin individually (i.e., tensored with identity matrices for the nuclear spin and the other electronic spin); for the dephasing noise, $L_{i}$ is only $\sigma_{z}$ for each electronic spin individually; for the depolarized noise, $L_{i}$ are $\sigma_{x}$, $\sigma_{y}$, $\sigma_{z}$ for each electronic spin individually.

Firstly, let us examine the effect of uncorrelated amplitude damping noise, with the numerical results shown in Fig. A5.
And we can see from Fig. A5(a) that when $\Gamma=0.1k$, the QFI of $30\%$ weaker and stronger fields are almost not changed compared with that of geomagnetic field. Moreover, a weak resonant oscillating field $\mathrm{B}_{\mathrm{rf}}=150$nT, which is perpendicular to Earth's magnetic field, can highly reduce the QFI with $\mathrm{B}_{\mathrm{0}}=46\mu$T, and the percent decrease of QFI, i.e., $\Delta\mathrm{QFI}/\mathrm{QFI}$, can be larger than $80\%$ shown in the inset of Fig. A5(a), which is large enough to imply that a weak oscillating field can completely disrupt the birds. And for $\Gamma=k$ in Fig. A5(b), the $30\%$ stronger and weaker fields still have little influences on the value of QFI. Meanwhile, there still exists an obvious difference in the value of QFI with and without the oscillating field, with $\Delta\mathrm{QFI}/\mathrm{QFI}$ being larger than $50\%$ shown in the inset of Fig. A5(b). However, when $\Gamma=10k$ in Fig. A5(c), we can see that although the curves of QFI for different magnetic field intensities overlap completely, it would render the bird almost immune to the weak oscillating field, with $\Delta\mathrm{QFI}/\mathrm{QFI}$ being smaller than $10\%$ shown in the inset of Fig. A5(c), which can not account for the fact that a weak oscillating field can completely disrupt the avian compass. As a conclusion, the decoherence rate should be smaller than $10 k$ for this amplitude damping noise.

Next, we consider the effect of uncorrelated dephasing noise, with the numerical results shown in Fig. A6. From Fig. A6 we can see that the $30\%$ stronger and weaker fields compared with the geomagnetic field have almost no influences on the value of QFI for $\Gamma=0.1k,~k$ and $10k$. However, the effects of a resonant oscillating field on the value of QFI are different. Specifically, from Fig. A6(a), we can see that when $\Gamma=0.1k$, the QFI would be highly reduced when the oscillating field is applied, with $\Delta\mathrm{QFI}/\mathrm{QFI}$ being larger than $80\%$ shown in the inset of Fig. A6(a). And when $\Gamma=k$, the oscillating field is still able to reduce the value of QFI to some extent, especially when $\theta$ is small, $\Delta\mathrm{QFI}/\mathrm{QFI}$ can reach approximately $90\%$ shown in the inset of Fig. A6(b). But when $\Gamma=10k$, we can see from Fig. A6(c) that bird becomes quite immune to the weak oscillating field with $\Delta\mathrm{QFI}/\mathrm{QFI}$ being approximately equal to 0 for large $\theta$. Thus for this dephasing noise, the decoherence rate should be smaller than $10 k$.

\begin{widetext}
\begin{center}
\includegraphics[width=15cm]{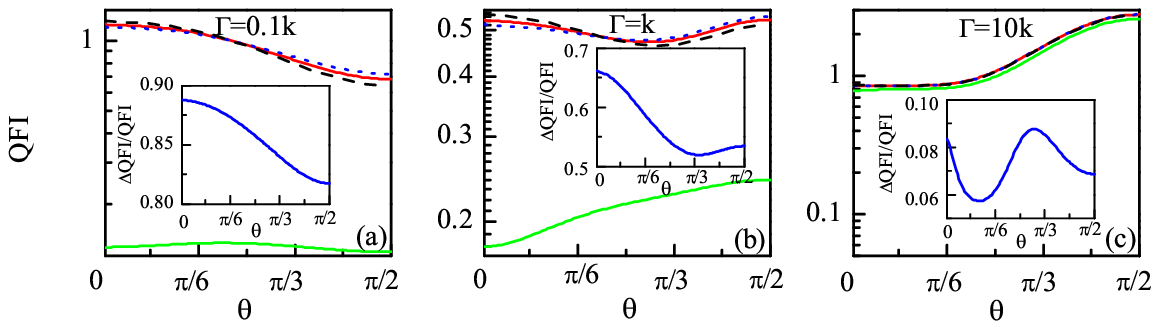}
\parbox{15cm}{\small{FIG. A5.} (Color online) The QFI as a function of the direction angle $\theta$ for the amplitude damping noise for (a) $\Gamma=0.1k$, (b) $\Gamma=k$, and (c) $\Gamma=10k$ without the oscillating field ($\mathrm{B}_{\mathrm{0}}=46\mu$T (red solid line),
$\mathrm{B}_{\mathrm{0}}=59.8\mu$T (black dashed line), and $\mathrm{B}_{\mathrm{0}}=32.2\mu$T (blue dotted line)), and with the oscillating field $\mathrm{B}_{\mathrm{rf}}=150$nT and $\mathrm{B}_{\mathrm{0}}=46\mu$T (green dash dotted line). The insets show the corresponding $\Delta\mathrm{QFI}/\mathrm{QFI}$ with the oscillating field. $A_{z} = 6\gamma\times46\mu$T, $A_{x}=A_{y}=0$, $k=10^4s^{-1}$.}
\end{center}
\end{widetext}

\begin{widetext}
\begin{center}
\includegraphics[width=15cm]{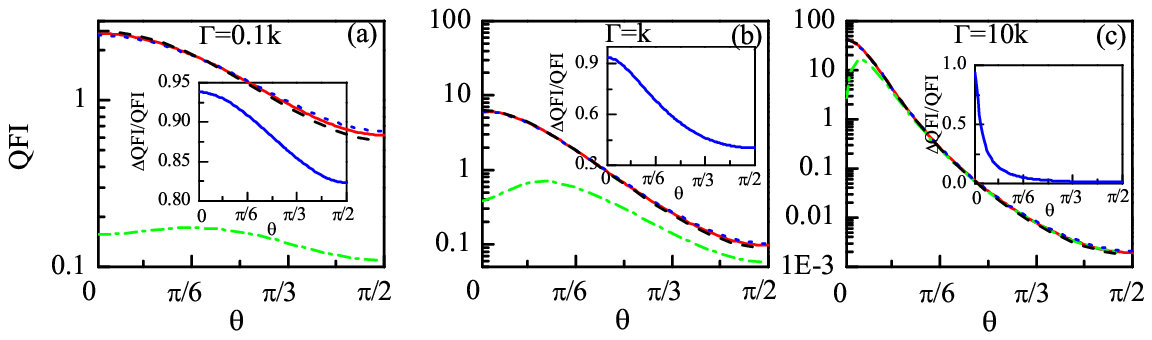}
\parbox{15cm}{\small{FIG. A6.} (Color online) The QFI as a function of the direction angle $\theta$ for the dephasing noise for (a) $\Gamma=0.1k$, (b) $\Gamma=k$, and (c) $\Gamma=10k$ without the oscillating field ($\mathrm{B}_{\mathrm{0}}=46\mu$T (red solid line), $\mathrm{B}_{\mathrm{0}}=59.8\mu$T (black dashed line), and $\mathrm{B}_{\mathrm{0}}=32.2\mu$T (blue dotted line)), and with the oscillating field $\mathrm{B}_{\mathrm{rf}}=150$nT and $\mathrm{B}_{\mathrm{0}}=46\mu$T (green dash dotted line). The insets show the corresponding $\Delta\mathrm{QFI}/\mathrm{QFI}$ with the oscillating field. $A_{z} = 6\gamma\times46\mu$T, $A_{x}=A_{y}=0$, $k=10^4s^{-1}$.}
\end{center}
\end{widetext}

Finally, we consider the effect of uncorrelated depolarized noise, with the numerical results shown in Fig. A7. From Fig. A7 we can see that when $\Gamma=0.1k$, the QFI of $30\%$ weaker and stronger fields are almost unchanged compared with that of geomagnetic field, and the difference in the value of QFI with and without the oscillating field is obvious, with the percent decrease $\Delta\mathrm{QFI}/\mathrm{QFI}$ being larger than $59\%$ shown in the inset of Fig. A7(a). However, when $\Gamma\geq k$, on the one hand, the value of QFI is significantly small despite of its insensitivity to the $30\%$ weaker and stronger fields. On the other hand, the oscillating field has almost no effect on the value of QFI, with $\Delta\mathrm{QFI}/\mathrm{QFI}$ being approximately $4\%$ shown in the inset of Fig. A7(b) or smaller than $0.1\%$ shown in the inset of Fig. A7(c). As a result, for this uncorrelated depolarized noise, the decoherence rate should be smaller than $1k$.

\begin{widetext}
\begin{center}
\includegraphics[width=15cm]{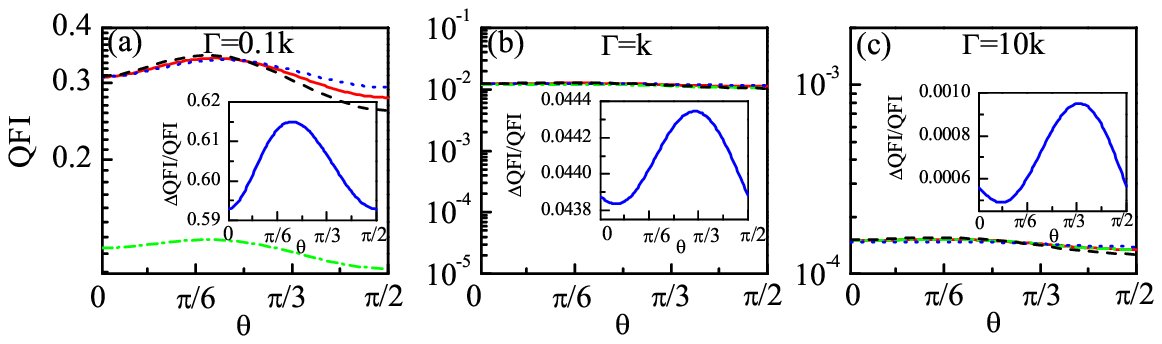}
\parbox{15cm}{\small{FIG. A7.} (Color online) The QFI as a function of the direction angle $\theta$ for the depolarized noise for (a) $\Gamma=0.1k$, (b) $\Gamma=k$, and (c) $\Gamma=10k$ without the oscillating field ($\mathrm{B}_{\mathrm{0}}=46\mu$T (red solid line), $\mathrm{B}_{\mathrm{0}}=59.8\mu$T (black dashed line), and $\mathrm{B}_{\mathrm{0}}=32.2\mu$T (blue dotted line)), and with the oscillating field $\mathrm{B}_{\mathrm{rf}}=150$nT and $\mathrm{B}_{\mathrm{0}}=46\mu$T (green dash dotted line). The insets show the corresponding $\Delta\mathrm{QFI}/\mathrm{QFI}$ with the oscillating field. $A_{z} = 6\gamma\times46\mu$T, $A_{x}=A_{y}=0$, $k=10^4s^{-1}$.}
\end{center}
\end{widetext}

\end{document}